\documentclass[showpacs,preprintnumbers,amsmath,amssymb,prl,twocolumn,superscriptaddress]{revtex4}

\usepackage{dcolumn}% Align table columns on decimal point
\usepackage{bm}
\usepackage{epsfig}% bold math
\begin{document}
%\draft
%\input{mathmac}
% Formula MACROS enter here #####################################################

%
% End of formula MACRO list ######################################################
%
%%\twocolumn[\hsize\textwidth\columnwidth\hsize
%%\csname@twocolumnfalse\endcsname
%

%
%%]
%
% Formula MACROS enter here #####################################################
\noindent {{\bf Reply to 'Difficulty in the Fermi-Liquid-Based
Theory for the In-Plane Magnetic Anisotropy in Untwinned High-$T_c$
Superconductors'}}

%\affiliation{$^1$ Max--Planck--Institut f\"ur Physik komplexer
%Systeme, Noethnitzer Strasse 38, D--01187 Dresden, Germany and \\
%Institute f\"ur Theoretische and Mathematische Physik, Technical
%University Braunschweig, D--38106 Braunschweig, Germany}
%\address{$^2$Max--Planck--Institut f\"ur Festk\"orperforschung,
%Heisenbergstrasse 1, D--70569 Stuttgart, Germany}
%\date{\today}

%\setcounter{equation}{0}
%\setcounter{figure}{0}
%%{\it Introduction}

%In a preceeding comment \cite{li}, Zhao and Li argued that the use
%of a negative value of the orthorhombic parameter $\delta_0$ that is
%necessary for explaining inelastic neutron scattering (INS)
%experiments, is inconsistent with ARPES data on untwinned YBCO which
%would invalidate a one-band Fermi-liquid theory.

We show that although our original paper contains some unprecise
statements there is neither a contradiction between ARPES and INS in
assuming a $\delta_0 < 0$, nor significant physical errors that
affect our results. In particular, we did never {\it proved}
$\delta_0
> 0$; instead, $\delta_0 = -0.03$ had been already used in Fig. 4(b)
of Ref. \cite{manske}.

Let us first note, that indeed a positive parameter $\delta_0$ would
be consistent with the result of a simple quantum-chemical
calculation yielding $t_a / t_b \sim \left( b / a \right)^{4}$.
%
%However, an {\it ab-initio} calculation that takes the presence of
%CuO chains between the CuO$_2$-planes into account (running along
%the $b$-axis), yields $t_b = -558$meV and $t_a = -537$meV and thus
%$\mid t_b \mid > \mid t_a\mid$ for the nearest-neighbor hopping (see
%Fig. 1 in Ref. \cite{okaandliechtenstein} and also Ref.
%\cite{andersen}.)
%
On the other hand, the LDA Fermi surface (FS) is more complicated
than the one-band model used by us. In particular, the saddlepoints
are not at $(\pm\pi,0)$ and $(0,\pm\pi)$, but bifurcated
considerably away from this \cite{okaandliechtenstein}. Furthermore,
we have neglected both the bonding plane band and the chain band. A
recent work by O.K. Andersen and co-workers calculating the {\it
downfolded} plane bands reproduce the LDA result in detail, but in a
non-trivial way: the nearest-neighbor hopping amplitude is {\it
smaller} along the $x$- than along the $y$-direction, and the {\it
opposite} is true for longer-ranged hoppings \cite{andersen}. Thus,
using the notation of Ref. [\protect\onlinecite{schnyder}], one
finds $|t_{1x}| < |t_{1y}|$.
Because the first hopping matrix element is most important, we have
used $\delta_0 < 0$ in Fig. 4(b) of Ref. \cite{manske} in order to
explain the anisotropy in the inelastic neutron scattering (INS)
data by Hinkov {\it et al.} \cite{hinkov}. The result $|t_{1x}| <
|t_{1y}|$ makes it necessary to re-visit early ARPES data on
untwinned YBa$_2$Cu$_3$O$_{6+x}$ (YBCO) \cite{schabel}.
%
%Note, recent ARPES measurements on untwinned YBCO agree with LDA
%predictions \cite{borisenko}.
%
%%
%\begin{figure}[h]
%\centerline{\psfig{file=YBCOPlaneBandsHOP.eps,width=4cm,angle=0}}
%\caption{(color online). LDA calculation for the in-plane hopping
%matrix elements in orthorhombic YBCO. Despite the fact that $b > a$,
%one finds $\mid t_b \mid > \mid t_a\mid$. Taken from Ref.
%\protect\onlinecite{andersen}}\label{replyfig1}
%\end{figure}
%%
%
%
\begin{figure}[h]
\centerline{\psfig{file=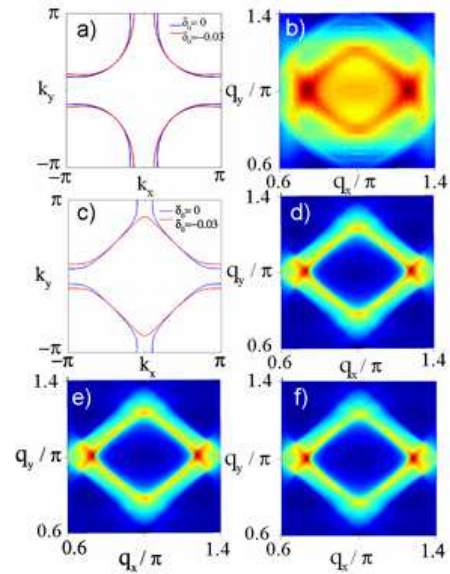,width=6.0cm}} \caption{(color
online). Comparison of the calculated INS intensity (b) and (d) for
two different Fermi surface topologies (a) and (c) calculated for
two different tight-binding parameters (tb2, tb3 from Ref.
\protect\cite{norman}). Different types of superconducting gaps are
compared in (e) $\Delta_k = \Delta_0^d \left(\cos k_x - \cos
k_y\right) / 2 + \Delta_0^s \left(\cos k_x + \cos k_y\right) / 2$
and (f) $\Delta_k = \Delta_0^d \left(\cos k_x - \cos k_y\right) / 2
+ \Delta_0^s$.} \label{replyfig2}
\end{figure}
However, in contrast to our Fig. 1 of Ref.
[\protect\onlinecite{erratum}] in untwinned YBCO, ARPES does not
observe a closing of the Fermi surface around $(0,\pm\pi)$. In order
to see whether a closed or opened Fermi surfaces would affect our
main conclusions we present in Fig. \ref{replyfig2} the result
obtained from our Fermi-liquid-based approach for $\delta_0 < 0$ and
two different tight-binding parameters. One clearly sees that,
although the FS topology changes, the calculated INS response
reveals basically the same result, i.e. two-dimensional and highly
anisotropic with two clear maxima along the $q_x$-direction (see
also Ref.\onlinecite{metzner}).
%Such a
%result has also been obtained in Ref.
%[\protect\onlinecite{schnyder}].
%
%\begin{figure}[h]
%\centerline{\psfig{file=replyfig2.eps,width=5cm}} \caption{(color
%online). Comparison of the calculated INS intensity at $\omega =
%35$meV for different types of superconducting gaps: (a) $\Delta_k =
%\frac{\Delta_0^d}{2}\left(\cos k_x - \cos k_y\right) +
%\frac{\Delta_0^s}{2}\left(\cos k_x + \cos k_y\right)$ and (b)
%$\Delta_k = \frac{\Delta_0^d}{2}\left(\cos k_x - \cos k_y\right) +
%\Delta_0^s$. Although both $s$-wave components strongly differ at
%$(\pm\pi,0)$ and $(0,\pm\pi)$, respectively, they yield almost the
%same INS intensity.} \label{replyfig3}
%\end{figure}
%

Finally, if Cooper-pairing is driven by a short-range interaction as
it is believed in high-$T_c$ cuprates, then the wave function $\psi
\sim \left(\cos k_x + \cos k_y\right)$ corresponds to the $s$-wave
component of the superconducting gap function. However, as Zhao and
Li correctly pointed out \cite{zhao}, this admixture of the $s$-wave
component to the original $d_{x^2-y^2}$-wave symmetry does {\it not}
yield the experimentally observed anisotropy of the superconducting
gap. In order to see whether this is a significant physical error we
present in (e) and (f) the calculated INS response for $\omega =
35$meV. One clearly sees that the results are nearly independent of
the particular $d+s$-wave gap representations.
%This
%issue has also been addressed in more detail in Ref.
%[\onlinecite{schnyder}].

In summary, for $\delta_0 =-0.03$ there exists no contradiction
between the calculated INS intensity and recent ARPES data. Thus,
our Fermi-liquid-based theory still provides an alternative approach
to the stripe scenario in order to explain highly anisotropic INS
data.
%
%Moreover, the exact choice of the $s$-wave component of the
%superconducting order parameter does not affect the spin
%susceptibility results for $\omega = 35$meV.

We thank V. Hinkov, W. Metzner, and H.Yamase for helpful discussions
and O.K. Andersen for providing us with LDA results prior to
publication.

\noindent I. Eremin$^1$ and D. Manske$^2$, $^1$Max--Planck--Institut
f\"ur Physik komplexer Systeme, N\"othnitzer Strasse 38, D--01187
Dresden and TU Braunschweig, D--38106 Braunschweig;
$^2$Max--Planck--Institut f\"ur Festk\"orperforschung,
Heisenbergstrasse 1, D--70569 Stuttgart, Germany

\vspace{-1.5ex} %sorry

\end{document}